\documentclass[onecolumn,useAMS,usenatbib]{mnras}
\usepackage{graphicx}
\usepackage{subfigure}
\usepackage{xcolor}
\usepackage{mathtext,bbm,amsmath,amsfonts,amssymb,indentfirst,syntonly,graphicx}
\usepackage{mathtools}
\usepackage{slashbox}
\usepackage[english]{babel}
\usepackage{calc}
\usepackage{tikz}
\usepackage[T1]{fontenc}
\usepackage{ae,aecompl}
\usepackage{latexsym,graphicx,bbm}
\usepackage{enumerate}
\usepackage{multirow}
\usepackage{hyperref}

\title[Statistical similarity between SGR and FRB]{Statistical similarity between soft gamma repeaters and repeating fast radio bursts }
\author[Y. Sang and H.-N. Lin]
{ Yu Sang$^{1, 2}$, Hai-Nan Lin$^{3}$\thanks{Corresponding author: linhn@cqu.edu.cn}\\
$^{1}$Center for Gravitation and Cosmology, College of Physical Science and Technology, Yangzhou University, Yangzhou 225009, China\\
$^{2}$Shanghai Frontier Science Research Center for Gravitational Wave Detection, School of Aeronautics and Astronautics, \\
Shanghai Jiao Tong University, Shanghai 200240, China\\
$^{3}$Department of Physics, Chongqing University, Chongqing 401331, China\\}
\begin{document}

\date{Accepted xxxx; Received xxxx; in original form xxxx}

\pagerange{\pageref{firstpage}--\pageref{lastpage}} \pubyear{2021}

\maketitle

\label{firstpage}

\begin{abstract}
  We study the statistical properties of the soft gamma repeater SGR 1935+2154. We find that the cumulative distributions of duration, waiting time, fluence and flux can be well fitted by bent power law. In addition, the probability density functions of fluctuations of duration, waiting time, fluence and flux well follow the Tsallis $q$-Gaussian distribution. The $q$ values keep steady for different temporal scale intervals, indicating a scale-invariant structure of the bursts. Those features are very similar to the property of the repeating fast radio burst FRB 121102, indicating the underlying association between the origins of soft gamma repeaters and repeating fast radio bursts.
\end{abstract}


\section{Introduction}\label{sec:introduction}

Soft gamma repeaters (SGRs) are high energy transients with persistent emissions of hard X-ray and  soft $\gamma$-ray bursts, with typical durations of $\thicksim$ 100 ms and peak luminosity up to $\thicksim10^{42}~\textrm{erg s}^{-1}$ \citep{Duncan1992,Kouveliotou1998,Kouveliotou1999,Thompson2002}. As a popular model to explain the SGRs, magnetar is a kind of isolated neutron stars with extremely strong magnetic fields of $\thicksim10^{14}-10^{15}$ G, which power the persistent hard X-ray and soft gamma-ray emissions \citep{Mereghetti2008,Kaspi:2017fwg}. The rotational periods of magnetars are very long ($P \thicksim 2-12$ s), and increase with relatively large periods derivative ($\dot{P}\thicksim10^{-13}-10^{-10}~\textrm{s~s}^{-1}$).

As one of the most active and most widely studied SGRs, SGR 1935+2154 was discovered on 2014 July 5, when a short burst triggered the Burst Alert Telescope (BAT) onboard \emph{Swift} \citep{Stamatikos:2014}. The subsequent monitoring between July, 2014 and March, 2015 by \emph{Chandra} and \emph{XMM-Newton} measured its spin period to be $P = 3.24$ s and spin-down rate to be $\dot{P} = 1.43 \times 10^{-11}~\textrm{s~s}^{-1}$. This implies a magnetar-like dipolar magnetic field of $B \thicksim 2.2 \times10^{14}$ G at the equator, confirming it as a magnetar source \citep{Israel:2016rct}. Since the discovery of SGR 1935+2154, it has become one of the most recurring transient magnetars ever observed, exhibited burst active episodes almost annually \citep{Younes:2017xyt,Lin:2020mlk}. In particular, \citet{Lin:2020mlk} presented an investigation of 127 bursts from SGR 1935+2154 observed by the Gamma-ray Burst Monitor (GBM) onboard \emph{Fermi} and BAT onboard \emph{Swift} during its four active episodes in 2014, 2015 and 2016. After the detailed temporal and spectral analyses of the bursts, they found that SGR 1935+2154 emitted 3, 24, 42, and 54 bursts in 2014, 2015, May 2016 and June 2016, respectively. On 2020 April 27, SGR 1935+2154 entered its most prolific episode since its discovery, emitting hundreds of X-ray bursts over a few minutes. Six hours after the activity onset, the burst storm of SGR 1935+2154 was observed by NICER telescope. 217 bursts were detected during the first 1120 seconds, corresponding to a burst rate of $>0.2$ bursts $\textrm{s}^{-1}$ \citep{Younes:2020hie}.

On April 28, a bright Fast Radio Burst (FRB 200428) was independently detected from the direction of SGR 1935+2154 by CHIME \citep{CHIMEFRB:2020abu} and STARE2 \citep{Bochenek:2020zxn} radio telescopes. Simultaneous to the radio burst, a magnetar-like burst from SGR 1935+2154 was detected by multiple hard X-ray telescopes, with a spectrum harder than previously observed bursts from the same source \citep{Mereghetti:2020unm,InsightHXMTTeam:2020dmu,Ridnaia:2020gcv,Tavani:2020adq,Younes:2020tac}. Following FRB 200428, SGR 1935+2154 has shown several millisecond radio bursts with three to six magnitudes dimmer than FRB 200428 \citep{Kirsten:2020yin}. The discovery of exceptional FRB--X-ray burst association supports magnetars as the potential of at least some extragalactic FRBs.

The magnetar association of FRBs motivate us to study the common properties shared by the two transients. Several works have been done to study this issue \citep{Chang:2017bnb,Lin:2019ldn,Wadiasingh:2019wea}. For example, the power law distribution of the energy have been found in both SGRs \citep{Cheng1996,Gogus1999,Gogus2000,Chang:2017bnb,Cheng:2019ykn} and the repeating FRB 121102 \citep{Wang:2016lhy,Wang:2017agh,Wang:2019sio,Lin:2019ldn}. Besides the power law property, the scale invariance of the size fluctuations has also been investigated in SGR \citep{Chang:2017bnb} and repeating FRB 121102 \citep{Lin:2019ldn}. The power law distributions and scale-invariant structure of the energy fluctuations are predicted by self-organized criticality (SOC) systems \citep{Bak:1987xua,aschwanden2011self}, which occurs in many natural systems  exhibiting nonlinear energy dissipation. \citet{Cheng1996} investigated the statistical properties of 111 bursts from SGR 1806--20, and found that the cumulative energy distribution of bursts is similar to the well-known earthquake Gutenberg--Richter power law. They also found the similarity of waiting time and duration distributions between SGRs and earthquakes. The properties of scale invariance of earthquakes and SGRs were found by \citet{Wang:2015nsl} and \citet{Chang:2017bnb}, respectively. These studies on the statistical properties of SGRs and earthquakes support the idea that SGRs may be powered by crustquakes of magnetars.

Since SGR 1935+2154 is so far the only magnetar associated with a FRB, the study on its statistical properties is helpful for revealing the mystery of the origin of FRBs. In this paper, we study the statistical properties of SGR 1935+2154 in details, including the distribution of bursts and fluctuations. In Section \ref{sec:CDF}, we study the cumulative distributions of  duration, waiting time, fluence and flux of SGR 1935+2154. In Section \ref{sec:fluctuations}, we investigate the probability distribution functions of fluctuations of duration, waiting time, fluence and flux. Finally, in Section \ref{sec:conclusions}, we give the discussion and conclusions.

\section{the cumulative distribution function of SGR 1935+2154}\label{sec:CDF}

The first sample used in our paper is the 217 bursts detected on 2020 April 28 from the \emph{NICER} observations \citep{Younes:2020hie}. The NICER sample contains the burst start time ($T_{\rm st}$), end time ($T_{\rm et}$), burst duration ($T_{90}$) and flux. We study the distribution functions of $T_{90}$, waiting time (WT), fluence and flux using NICER sample. The $T_{90}$ burst duration is defined as the time interval during which the cumulative burst fluence rises from 5\% to 95\% \citep{Kouveliotou:1993yx}. The WT between successive bursts is defined as $T_{{\rm st}, i+1  } - T_{{\rm et}, i}$, where $i=1,2,3...$ is the burst number in temporal order. The burst fluence is calculated by multiplying the time-averaged flux by $T_{90}$. The second sample is the 112 bursts observed by \emph{Fermi}/GBM during the source's four active episodes from 2014 to 2016 \citep{Lin:2020mlk}. The GBM sample contains the burst start time ($T_{\rm burst }$), duration ($T_{90}$) and fluence, but no flux data is available. We study the distribution functions of $T_{90}$, WT and fluence using GBM sample. We calculate WT by the difference of $T_{\rm burst}$ of successive bursts, ${\rm WT}_{i} = T_{{\rm burst},i+1 } - T_{{\rm burst}, i }$. Since GBM sample is observed in four active episodes, we discard the WT between the last burst of the former active episode and the first burst of the later active episode.

\begin{figure}
 \centering
 \includegraphics[width=0.4\textwidth]{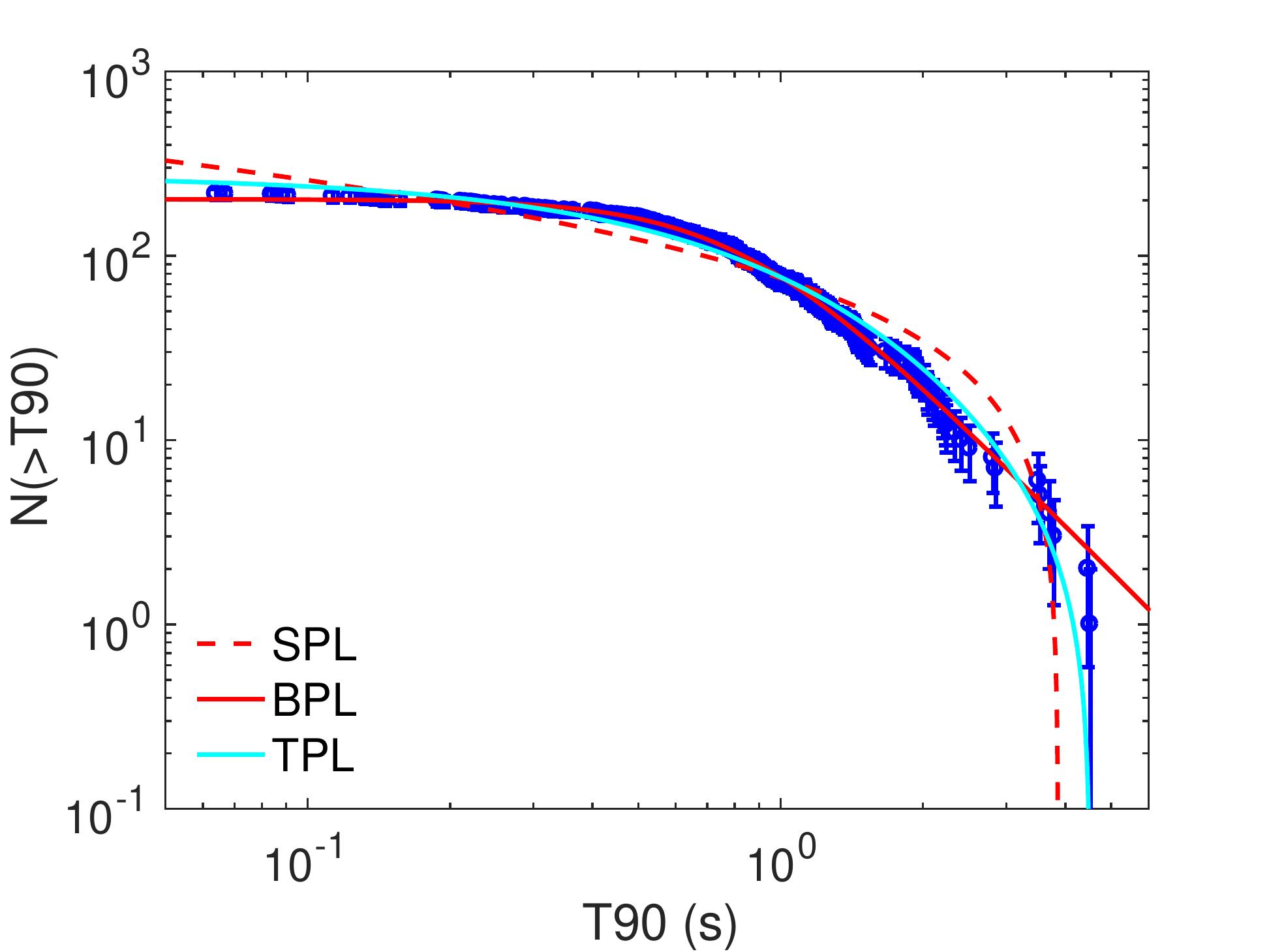}
 \includegraphics[width=0.4\textwidth]{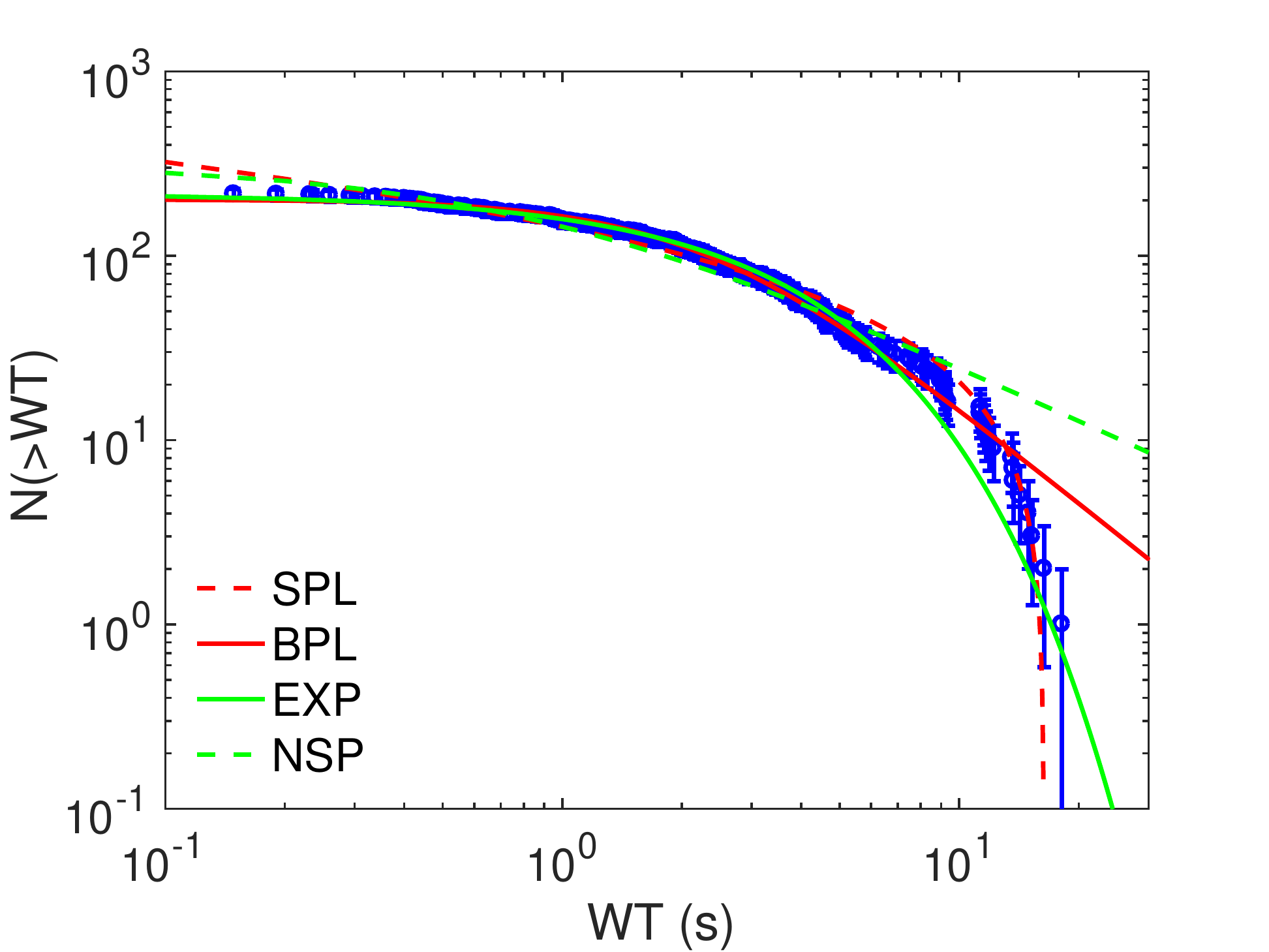}
 \includegraphics[width=0.4\textwidth]{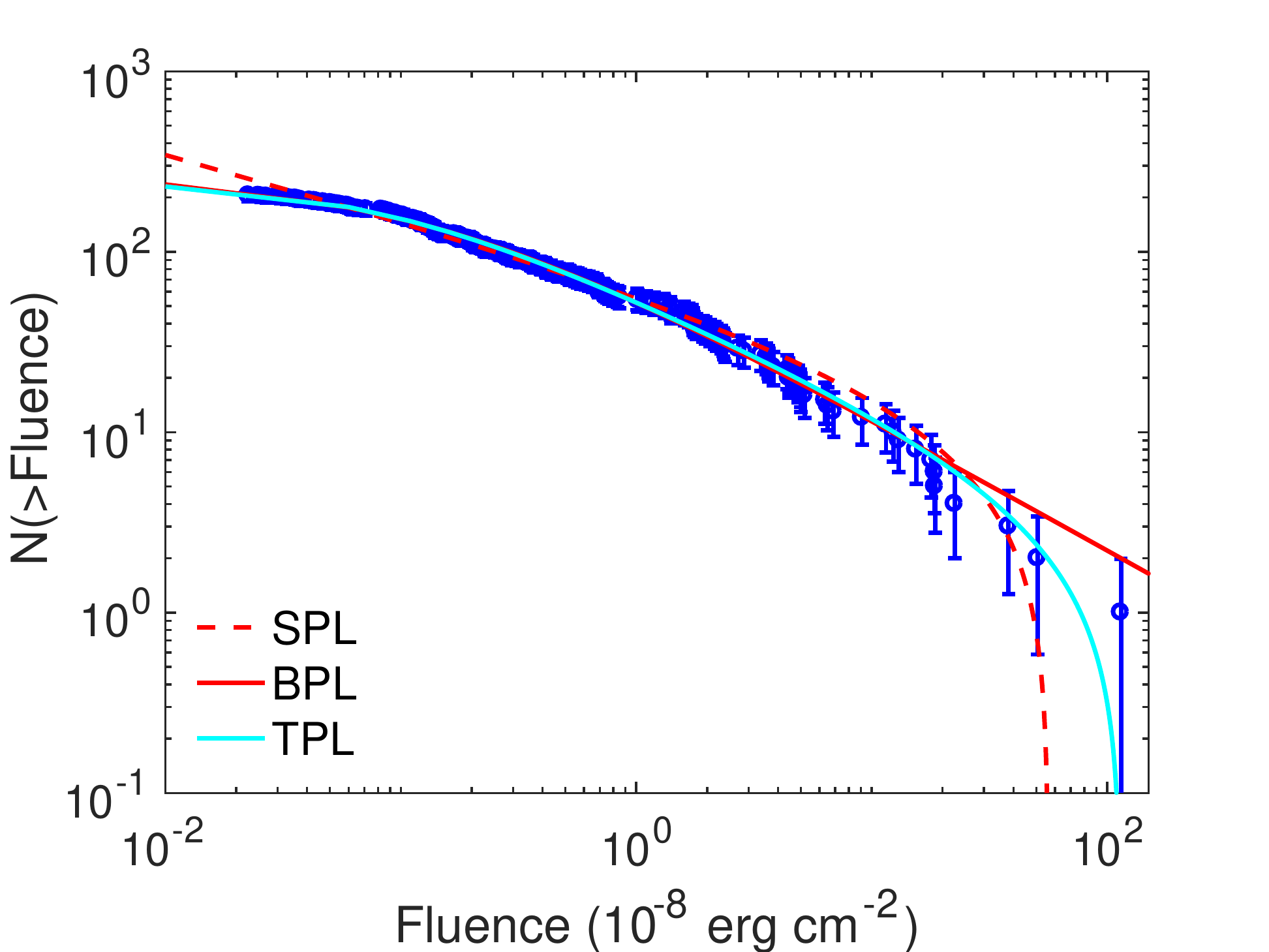}
 \includegraphics[width=0.4\textwidth]{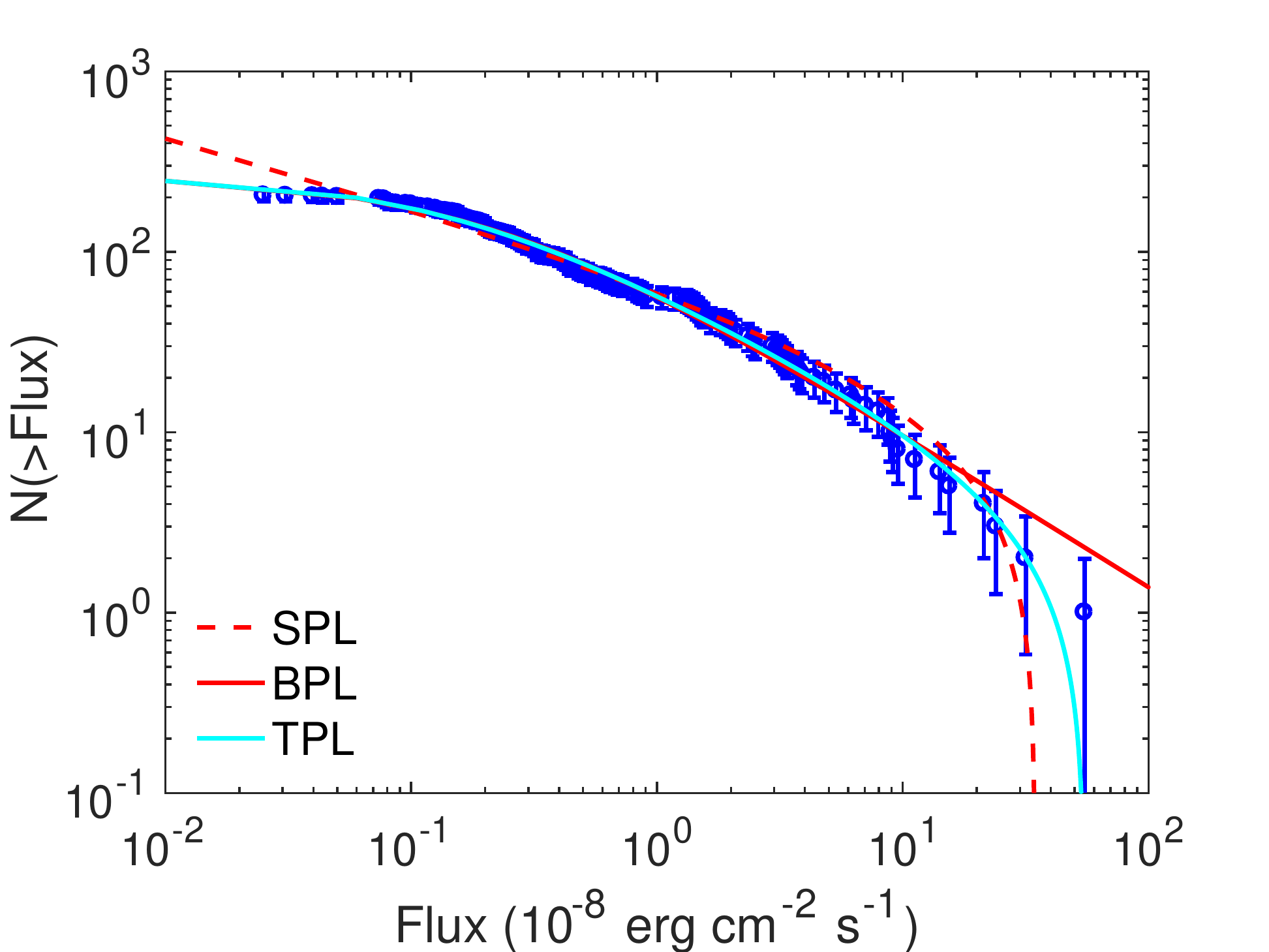}
 \caption{The CDFs of $T_{90}$, WT, fluence and flux for NICER sample. The red dashed, red solid and cyan solid lines are the best-fitting curves to the SPL, BPL and TPL models, respectively. The green solid and green dashed lines are the best-fitting curves to the EXP and NSP models, respectively.}
 \label{fig:cdf_NICER}
\end{figure}

Since the burst numbers in both samples are not large enough, instead of using differential distribution, we use cumulative distribution in the statistical analysis. The cumulative distribution functions (CDFs) of $T_{90}$, WT, fluence and flux for NICER sample are shown in Figure \ref{fig:cdf_NICER}. We first try to fit the CDFs using the simple power law (SPL) model. The power law distribution of duration is a prediction of SOC theory \citep{aschwanden2011self}. SOC theory also predicts the power law distribution of energy and waiting time. For example, \citet{Wang:2016lhy} showed the power law distribution of duration and fluence in the repeating FRB 121102, and \citet{Cheng:2019ykn} showed the power law distribution of duration and fluence in SGRs.
The SPL model is given by
\begin{equation}
  N(>x) = A (x^{-\alpha}-x_c^{-\alpha}), ~~~ x<x_c,
\end{equation}
where $x_c$ is the cut-off value above which $N(>x_c)=0$, $\alpha$ is the power-law index and $A$ is a normalization constant. We calculate the best-fitting parameters by minimizing the $\chi^2$,
\begin{equation}
  \chi^2=\sum_i\frac{[N_i-N(>x_i)]^2}{\sigma_i^2},
\end{equation}
where the uncertainty of data point is taken as $\sigma_i=\sqrt{N_i}$. For the NICER sample, the best-fitting parameters are listed in Table \ref{tab:para_NICER}. The power law indexes $\alpha$ of $T_{90}$, WT, fluence and flux are $0.19\pm0.03$, $0.16\pm0.02$, $0.35\pm0.01$ and $0.37\pm0.01$, respectively. The cut-off values $x_c$ of $T_{90}$, WT, fluence and flux are $3.87\pm0.10 ~{\rm s}$, $16.36\pm0.26 ~{\rm s}$, $(56.42\pm4.93) \times 10^{-8} ~{\rm erg ~ cm^{-2}}$ and $(34.64\pm3.33)\times 10^{-8} ~{\rm erg ~ cm^{-2}~ s^{-1}}$, respectively.
In Figure \ref{fig:cdf_NICER}, the red dashed lines are the best-fitting curves to the SPL model. The SPL model couldn't fit the data well, especially at left and right ends. The model predicts a larger value than observations at the left end, and predicts a sharp drop at the right end.

\begin{table}
  \centering
  \caption{The best-fitting parameters to the SPL, BPL, TPL, EXP and NSP models for the NICER sample. The units of $x_c$ ($x_b$ and $x_0$) for $T_{90}$, WT, Fluence and Flux are ${\rm s}$, ${\rm s}$, $10^{-8} ~{\rm erg ~ cm^{-2}}$ and $10^{-8} ~{\rm erg ~ cm^{-2}~ s^{-1}}$, respectively. The units of $\lambda$ and $\lambda_0$ are ${\rm s}^{-1}$.}
  \label{tab:para_NICER}
  \begin{tabular}{cccccc}
  \hline
  model & & $T_{90}$& WT & Fluence & Flux  \\
  \hline
  & $\alpha$ & $0.19\pm0.03$ & $0.16\pm0.02$ & $0.35\pm0.01$ & $0.37\pm0.01$ \\
  SPL & $x_c$ & $3.87\pm0.10$ & $16.36\pm0.26$ & $56.42\pm4.93$ & $34.64\pm3.33$ \\
  & $\chi^2_{\rm red}$ & $5.62$ & $1.35$ & $0.85$ & $1.62$ \\
  \hline
  & $\beta$ & $2.58\pm0.02$ & $1.74\pm0.02$ & $0.73\pm0.01$ & $0.85\pm0.02$ \\
  BPL & $x_b$ & $0.83\pm0.01$ & $2.29\pm0.03$ & $0.14\pm0.01$ & $0.21\pm0.02$ \\
  & $\chi^2_{\rm red}$ & $0.12$ & $0.38$ & $0.19$ & $0.44$ \\
  \hline
  & $\gamma$ & $8.56\pm2.68$ & \ldots & $1.57\pm0.01$ & $1.67\pm0.02$ \\
  TPL & $x_0$ & $5.62\pm2.21$ & \ldots & $0.08\pm0.01$ & $0.13\pm0.01$ \\
  & $\chi^2_{\rm red}$ & $1.15$ & \ldots & $0.15$ & $0.35$ \\
  \hline
  EXP & $\lambda$ & \ldots & $0.32\pm0.01$ & \ldots & \ldots \\
  & $\chi^2_{\rm red}$ & \ldots & 0.47 & \ldots & \ldots \\
  \hline
  NSP & $\lambda_0$ & \ldots & $1.19\pm0.07$ & \ldots & \ldots \\
  & $\chi^2_{\rm red}$ & \ldots & 3.93 & \ldots & \ldots \\
  \hline
  \end{tabular}
\end{table}

The data points in figure \ref{fig:cdf_NICER} show a flat tail at the left end, and show a power-law behavior at the right end. This motivates us to use the bent power law (BPL) model to fit the data. The BPL model is given by
\begin{equation}
  N(>x)=B\left[1+\left(\frac{x}{x_b}\right)^{\beta}\right]^{-1},
\end{equation}
where $x_b$ is the median value of $x$, i.e. $N(x>x_b)=B/2$, $B$ is the total number of data points, and $\beta$ is the power law index at $x\gg x_b$. The BPL model was first used to fit the power density spectra of gamma-ray bursts \citep{Guidorzi:2016ddt}. It was also used to fit the cumulative distributions of energy and waiting time of SGRs \citep{Chang:2017bnb} and FRBs \citep{Lin:2019ldn}. BPL is a phenomenological model, the underlying physical meaning is unclear at present. BPL is actually a piecewise power law model with a smooth connection. It has a flat tail at left end (power law index $p=0$), and has power-law behaviour at right end (power law index $p=\beta$). The asymptotic behavior of BPL model is consistent with the distribution of data points. For the NICER sample, the best-fitting parameters are listed in Table \ref{tab:para_NICER}. The power law indexes $\beta$ of $T_{90}$, WT, fluence and flux are $2.58\pm0.02$, $1.74\pm0.02$, $0.73\pm0.01$ and $0.85\pm0.02$, respectively. The median values $x_b$ of $T_{90}$, WT, fluence and flux are $0.83\pm0.01 ~{\rm s}$, $2.29\pm0.03~{\rm s}$,  $(0.14\pm0.01) \times 10^{-8} ~{\rm erg ~ cm^{-2}}$ and $(0.21\pm0.02)\times 10^{-8} ~{\rm erg ~ cm^{-2}~ s^{-1}}$, respectively. In Figure \ref{fig:cdf_NICER}, the red solid lines are the best-fitting curves to BPL model. From Table \ref{tab:para_NICER}, we see that BPL model has much smaller $\chi^2_{\rm red}$ value than SPL model, so BPL model is better than SPL model. From Figure \ref{fig:cdf_NICER} we can also see that SPL model couldn't fit the data well at left end, but BPL model fits the data well. In general, a good model should have $\chi^2_{\rm red}$ value equals to $\sim 1$. In our case, the $\chi^2_{\rm red}$ values for the BPL model are much smaller than 1. We interpret this by the overestimation of the uncertainty.

To compare our results with others, we also use the thresholded power-law (TPL) size distribution to fit the data. This model has been used to study the statistical properties of magnetar bursts and FRBs \citep{Cheng:2019ykn}. The differential distribution of TPL model is given by
\begin{equation}
   N(x)dx = n_0 (x_0 + x)^{- \gamma } dx,\quad x_1\leq x\leq x_2,
\end{equation}
where $x_0$ is a free parameter, $n_0$ is a normalization constant, $x_1$ and $x_2$ are the lower and upper cutoffs, respectively. The cumulative distribution of the TPL model is given by
\begin{equation}
   N( >x) =  \int_{x}^{x_2} N(x)dx  = \frac{n_0}{1-  \gamma } [(x_0 + x_2)^{1- \gamma } - (x_0 + x)^{1- \gamma } ].
\end{equation}
For the NICER sample, the best-fitting parameters are listed in Table \ref{tab:para_NICER}. The power law index $\gamma$ of $T_{90}$, fluence and flux are $8.56\pm2.68$, $1.57\pm0.01$ and $1.67\pm0.02$, respectively. The parameter $x_0$ of $T_{90}$,  fluence and flux are $5.62\pm2.21 ~{\rm s}$, $(0.08\pm0.01) \times 10^{-8} ~{\rm erg ~ cm^{-2}}$ and $(0.13\pm0.01)\times 10^{-8} ~{\rm erg ~ cm^{-2}~ s^{-1}}$, respectively. In Figure \ref{fig:cdf_NICER}, the cyan solid lines are the best-fitting curves to the TPL model. The $\chi^2_{\rm red}$ values for the SPL, BPL and TPL models are listed in Table \ref{tab:para_NICER} for comparison. From Figure \ref{fig:cdf_NICER}, we can see that at the left end, TPL model and BPL model fit the fluence and flux data equally well. At the right end, however, TPL model is a little better than BPL model\footnote{The difference between TPL model and BPL model is actually small at the right end, although from the figure the difference seems to be large. This is because the figure is plotted in logarithm scale.}. For $T_{90}$ data, TPL model fits the data much worse than BPL model.

\begin{figure}
 \centering
 \includegraphics[width=0.4\textwidth]{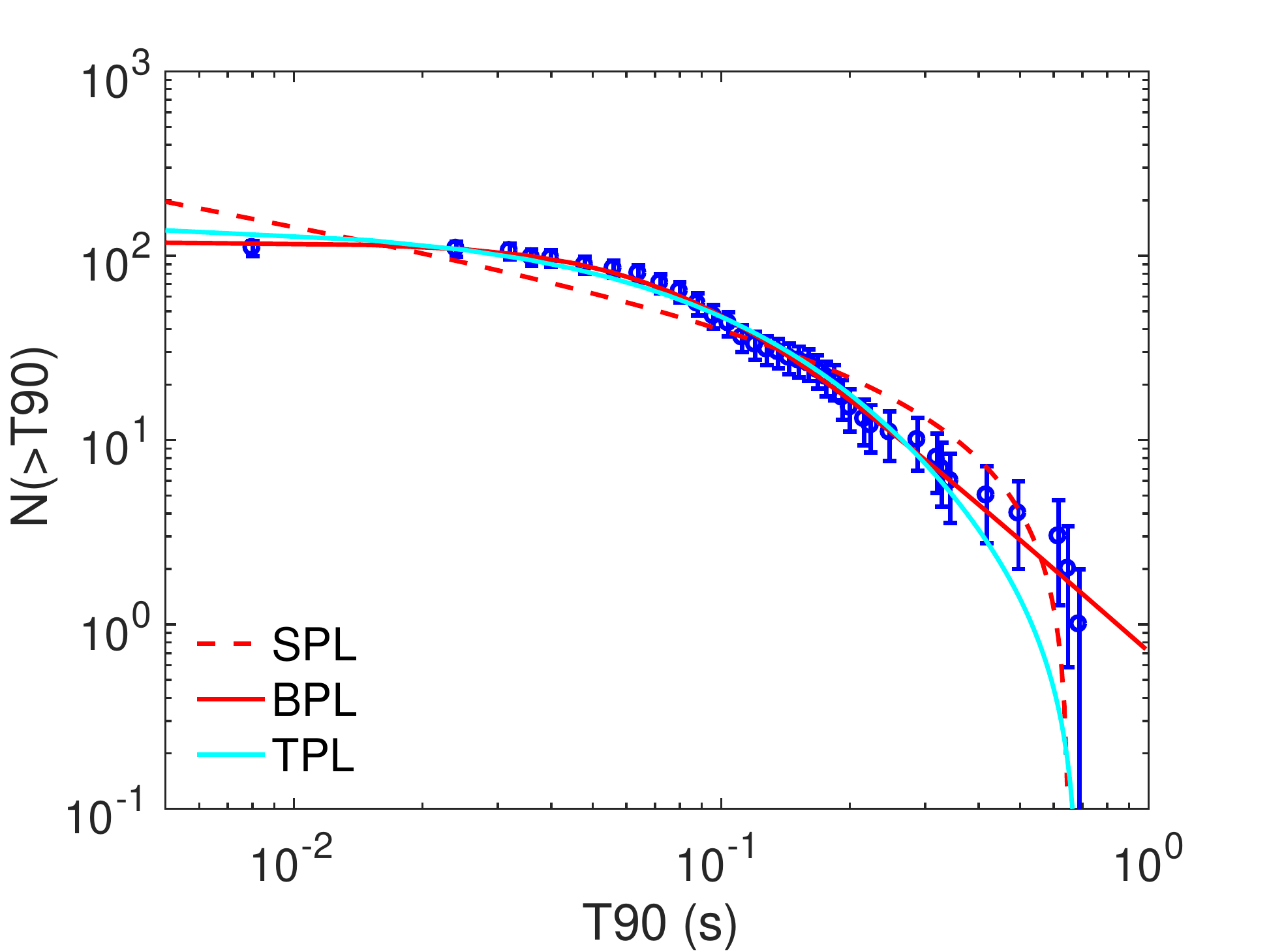}
 \includegraphics[width=0.4\textwidth]{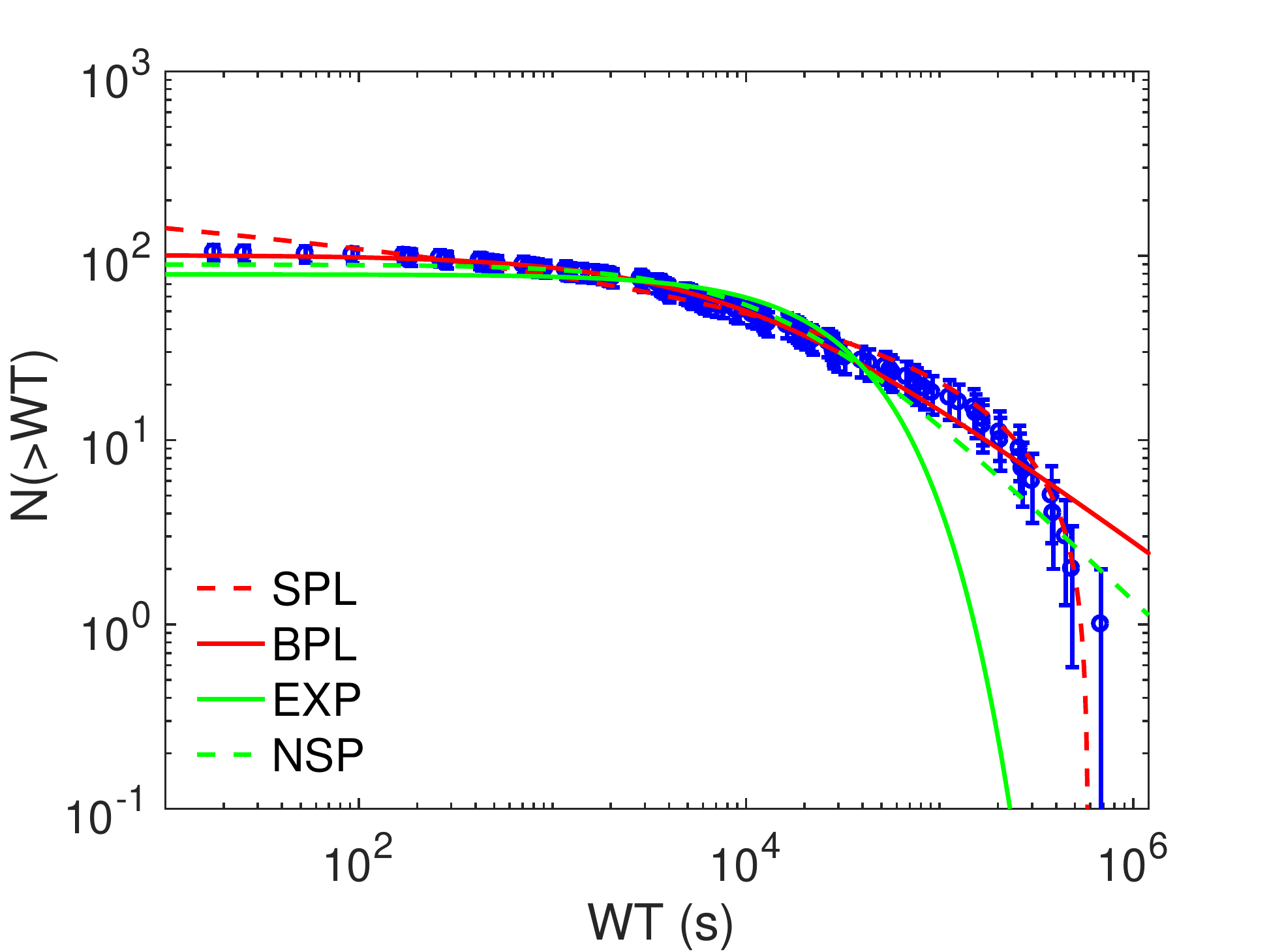}
 \includegraphics[width=0.4\textwidth]{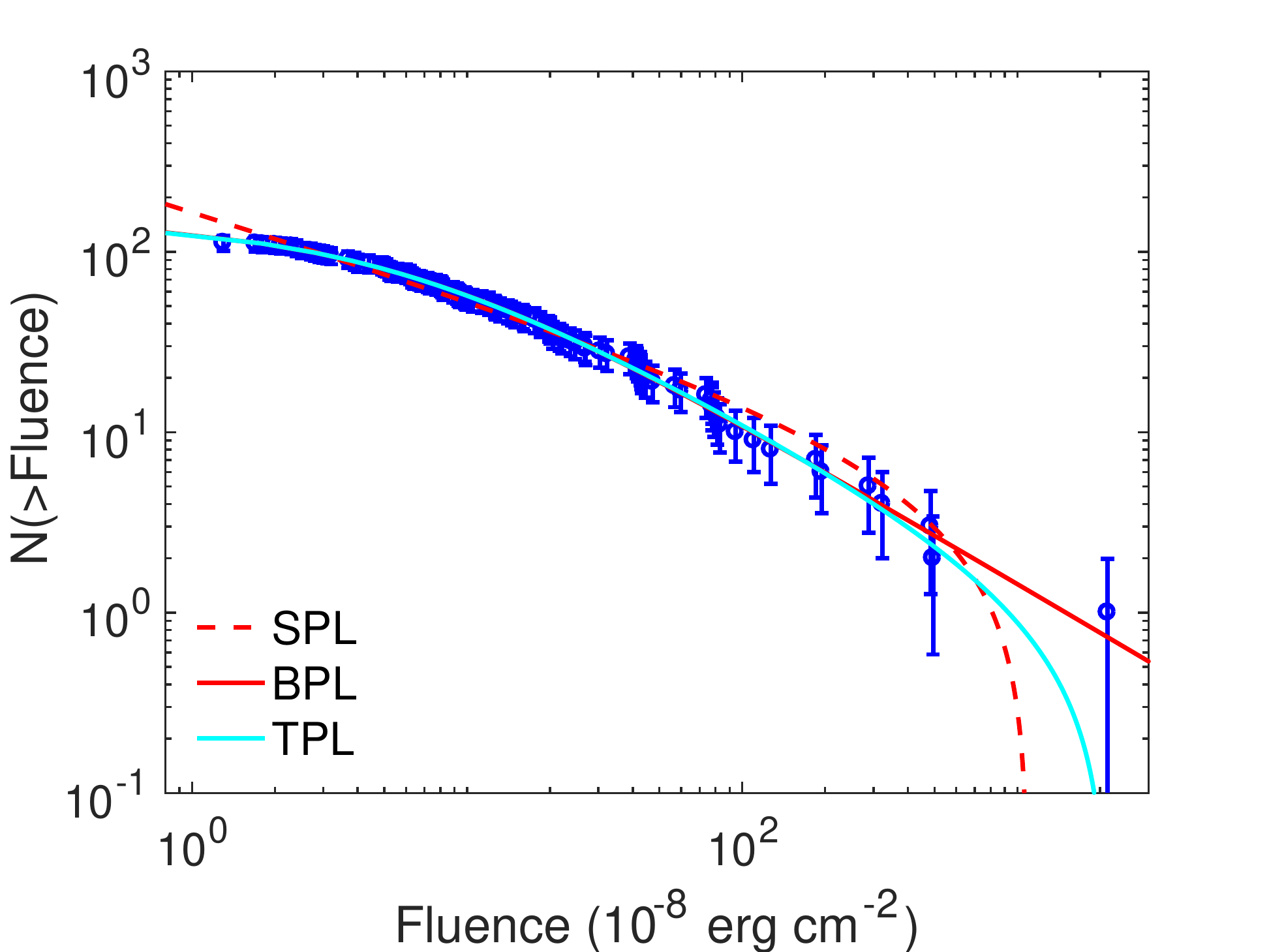}
 \caption{The CDFs of $T_{90}$, WT and fluence for GBM sample. The red dashed, red solid and cyan solid lines are the best-fitting curves to the SPL, BPL  and TPL models, respectively. The green solid and green dashed lines are the best-fitting curves to the EXP and NSP models, respectively.}
 \label{fig:cdf_Fermi}
\end{figure}

For the GBM sample, the CDFs of $T_{90}$, WT and fluence are shown by the blue error bars in Figure \ref{fig:cdf_Fermi}. We also use the SPL, BPL and TPL models to fit the CDFs of $T_{90}$, WT and fluence. The best-fitting parameters, together with the $\chi_{\rm red}^2$ values are listed in Table \ref{tab:para_Fermi} for comparison. The best-fitting curves are shown in the red dashed, red solid and cyan solid lines in Figure \ref{fig:cdf_Fermi} for the SPL, BPL and TPL models, respectively. For WT, BPL model is a little better than SPL model. The $\chi^2_{\rm red}$ value of the BPL model is smaller than that of the SPL model by a factor of $\sim2$. For $T_{90}$ and fluence, using BPL model instead of SPL model significantly improves the fits, with the $\chi^2_{\rm red}$ value reduced by a factor of more than 10. For the BPL model, the power law index $\beta$ of $T_{90}$, WT and fluence are $2.04\pm0.06$, $0.77\pm0.02$, and $0.90\pm0.01$, respectively. The median value $x_b$ of $T_{90}$, WT and fluence are $0.08\pm0.01 ~{\rm s}$, $(9.81\pm0.55)\times 10^3 ~{\rm s}$ and $(5.93\pm0.27) \times 10^{-8} ~{\rm erg ~ cm^{-2}}$, respectively.

\begin{table}
  \centering
  \caption{The best-fitting parameters to the SPL, BPL, TPL, EXP and NSP models for the GBM sample. The units of $x_c$ ($x_b$ and $x_0$) for $T_{90}$, WT and Fluence are ${\rm s}$, ${\rm s}$ and $10^{-8} ~{\rm erg ~ cm^{-2}}$, respectively. The units of $\lambda$ and $\lambda_0$ are ${\rm s}^{-1}$.}
  \label{tab:para_Fermi}
  \begin{tabular}{ccccc}
  \hline
  model & & T90& WT & Fluence   \\
  \hline
  & $\alpha$ & $0.36\pm0.07$ & $0.02\pm0.01$ & $0.46\pm0.01$  \\
  SPL & $x_c$ & $0.65\pm0.05$ & $(5.86\pm0.25)\times 10^5$ & $(1.09\pm0.18)\times 10^3$  \\
  & $\chi^2_{\rm red}$ & $3.44$ & $0.63$ & $0.58$ \\
  \hline
  & $\beta$ & $2.04\pm0.06$ & $0.77\pm0.02$ & $0.90\pm0.01$  \\
  BPL & $x_b$ & $0.08\pm0.01$ & $(9.81\pm0.55)\times 10^3 $ & $5.93\pm0.27$ \\
  & $\chi^2_{\rm red}$ & $0.18$ & $0.25$ & $0.05$  \\
  \hline
  & $\gamma$ & $6.72\pm3.11$ & \ldots & $1.81\pm0.02$  \\
  TPL & $x_0$ & $0.46\pm0.30$ & \ldots & $4.71\pm0.30$  \\
  & $\chi^2_{\rm red}$ & $0.68$ & \ldots & $0.06$  \\
  \hline
  EXP & $\lambda$ & \ldots & $(2.89\pm0.24)\times 10^{-5}$ & \ldots \\
  & $\chi^2_{\rm red}$ & \ldots & 2.87 & \ldots \\
  \hline
  NSP & $\lambda_0$ & \ldots & $(6.56\pm0.33)\times 10^{-5}$ & \ldots \\
  & $\chi^2_{\rm red}$ & \ldots & 0.64 & \ldots \\
  \hline
  \end{tabular}
\end{table}

If the bursts occur continuously and independently at a constant average occurrence rate, the waiting time is expected to follow the exponential (EXP) distribution \citep{aschwanden2011self}
\begin{equation}
   N(>\Delta t) \propto e^{-\lambda \Delta t},
\end{equation}
where $\lambda$ is the average occurrence rate. We use the EXP model to fit the CDF of WT of both NICER and GBM samples. The best-fitting parameters are $\lambda = 0.32\pm0.01  ~{\rm s}^{-1}$ ($\chi^2_{\rm red}=0.47$) for NICER sample and $\lambda = (2.89\pm0.24) \times 10^{-5} ~{\rm s}^{-1}$ ($\chi^2_{\rm red}=2.87$) for GBM sample. The NICER sample is observed during a very active bursts storm so its burst rate is much larger than the GBM sample. The green solid lines in Figure \ref{fig:cdf_NICER} and Figure \ref{fig:cdf_Fermi} are the best-fitting curves to the EXP model for NICER and GBM samples, respectively. For NICER sample, EXP model is better than SPL model, but is a little worse than BPL model. However, for GBM sample, the EXP model fails to fit data points.

If the burst occurrence rate is not a constant, the non-stationary Poisson (NSP) process should be used. For NSP process, the PDF of waiting time is given by \citep{aschwanden2011self}
\begin{equation}
   P(\Delta t) = \frac{\lambda_0}{(1+\lambda_0 \Delta t)^2}.
\end{equation}
This gives the CDF of waiting time
\begin{equation}
   N(>\Delta t) \propto \frac{ 1 }{1+\lambda_0 \Delta t}.
\end{equation}
We use the NSP model to fit the CDFs of WT for both NICER and GBM samples. The best-fitting parameters are $\lambda_0 = 1.19\pm0.07 ~{\rm s}^{-1}$ ($\chi^2_{\rm red}=3.93$) for NICER sample, and $\lambda_0 = (6.56\pm0.33) \times 10^{-5} ~{\rm s}^{-1}$ ($\chi^2_{\rm red}=0.64$) for GBM sample. The green dashed lines in Figure \ref{fig:cdf_NICER} and Figure \ref{fig:cdf_Fermi} are the best-fitting curves to NSP model for NICER and GBM samples, respectively. NSP model fits the GBM sample well, but fails to fit the NICER sample. In the upper-right panels of Figure \ref{fig:cdf_NICER} and Figure \ref{fig:cdf_Fermi}, we compare the SPL, BPL, EXP and NSP models. For both samples, the BPL model has the smallest $\chi^2_{\rm red}$ value ($\chi^2_{\rm red}=0.38$ for NICER sample and $\chi^2_{\rm red}=0.25$ for GBM sample), so BPL model is better than the other three models in fitting the WT data.

\section{Probability density functions of fluctuations}\label{sec:fluctuations}
In this section, we study the statistical properties of fluctuations of $T_{90}$, WT, fluence and flux. The fluctuation of a quantity $Q$  ($T_{90}$, WT, fluence, or flux) is given by $Z_n=Q_{i+n}-Q_i$, where $Q_i$ is the quantity of the $i$-th burst in temporal order, and $n$ is an integer denoting the temporal interval scale. $Z_n$ is usually rescaled as $z_n=Z_n/\sigma$, where $\sigma={\rm std}(Z_n)$ is the standard deviation of $Z_n$. Hence in this paper we study the statistical properties of the dimensionless fluctuation $z_n$. The fluctuations of bursts of SGR J1550-5418 and repeating FRB 121102 have been shown to follow the Tsallis $q$-Gaussian function \citep{Chang:2017bnb,Lin:2019ldn}. The Tsallis $q$-Gaussian function is defined as \citep{Tsallis:1987eu,Tsallis:1998ws}
\begin{equation}
  f(x)=\alpha[1-\beta(1-q)x^2]^{\frac{1}{1-q}},
\end{equation}
where $\alpha$ is the normalization factor, the parameters $\beta$ and $q$ control the width and sharpness of the peak, respectively.  The $q$-Gaussian function is a generalization of the Gaussian distribution. It has a much sharper peak at $x=0$ and fatter tails at left and right ends than the Gaussian function. The deviation from the Gaussian distribution is described by the parameter $q$. When $q\rightarrow 1$, the $q$-Gaussian function reduces to the Gaussian function with $\mu = 0$ and $\sigma=1/\sqrt{2\beta}$. Because of the limited number of data points, we use the CDF of $q$-Gaussian function to fit the fluctuations, which is given by
\begin{equation}
  F(x)=\int_{-\infty}^xf(x)dx.
\end{equation}
We consider the burst fluctuations in different temporal interval scale $n$ from 1 to 40.

\begin{figure}
 \centering
 \includegraphics[width=0.4\textwidth]{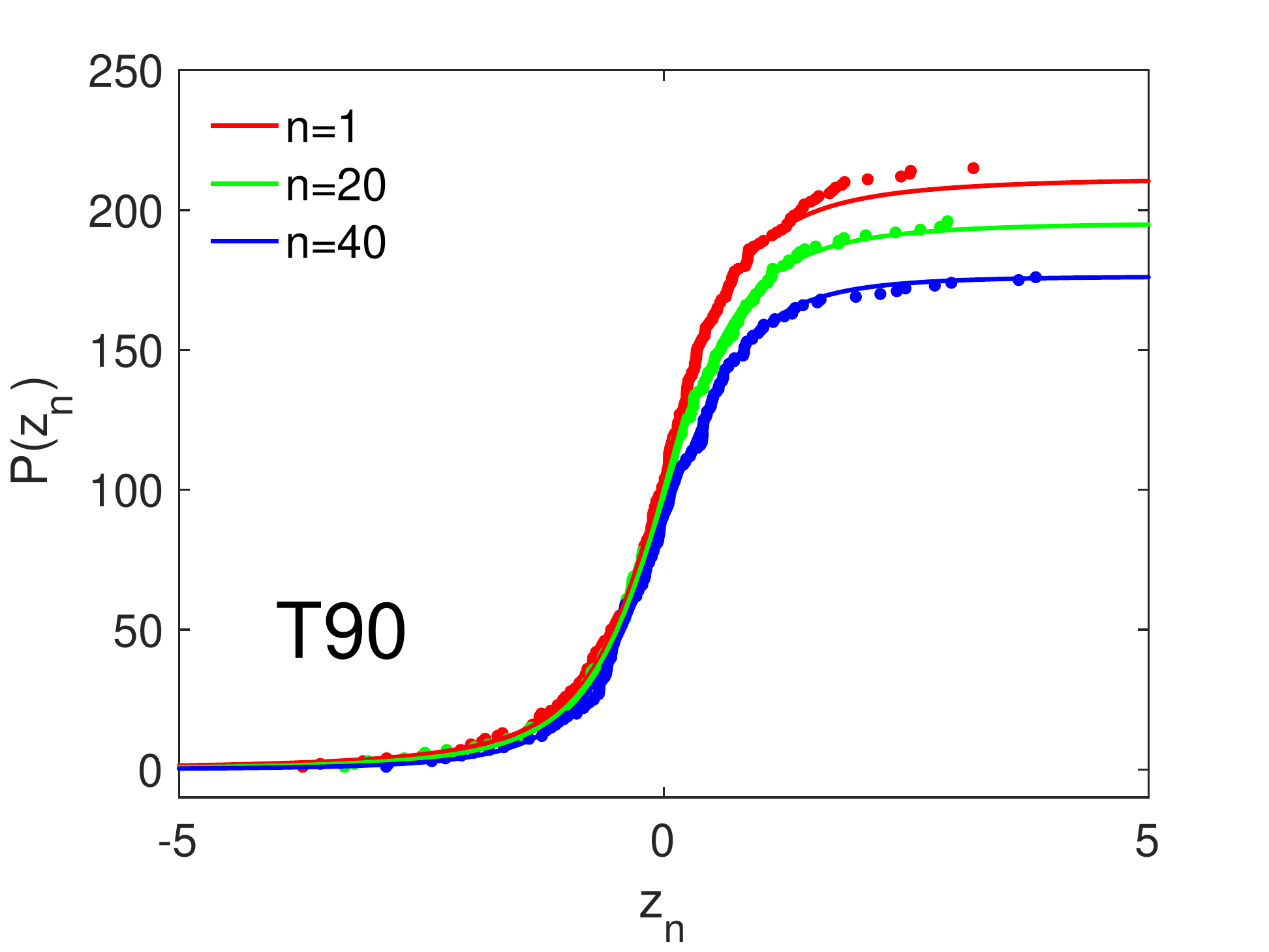}
 \includegraphics[width=0.4\textwidth]{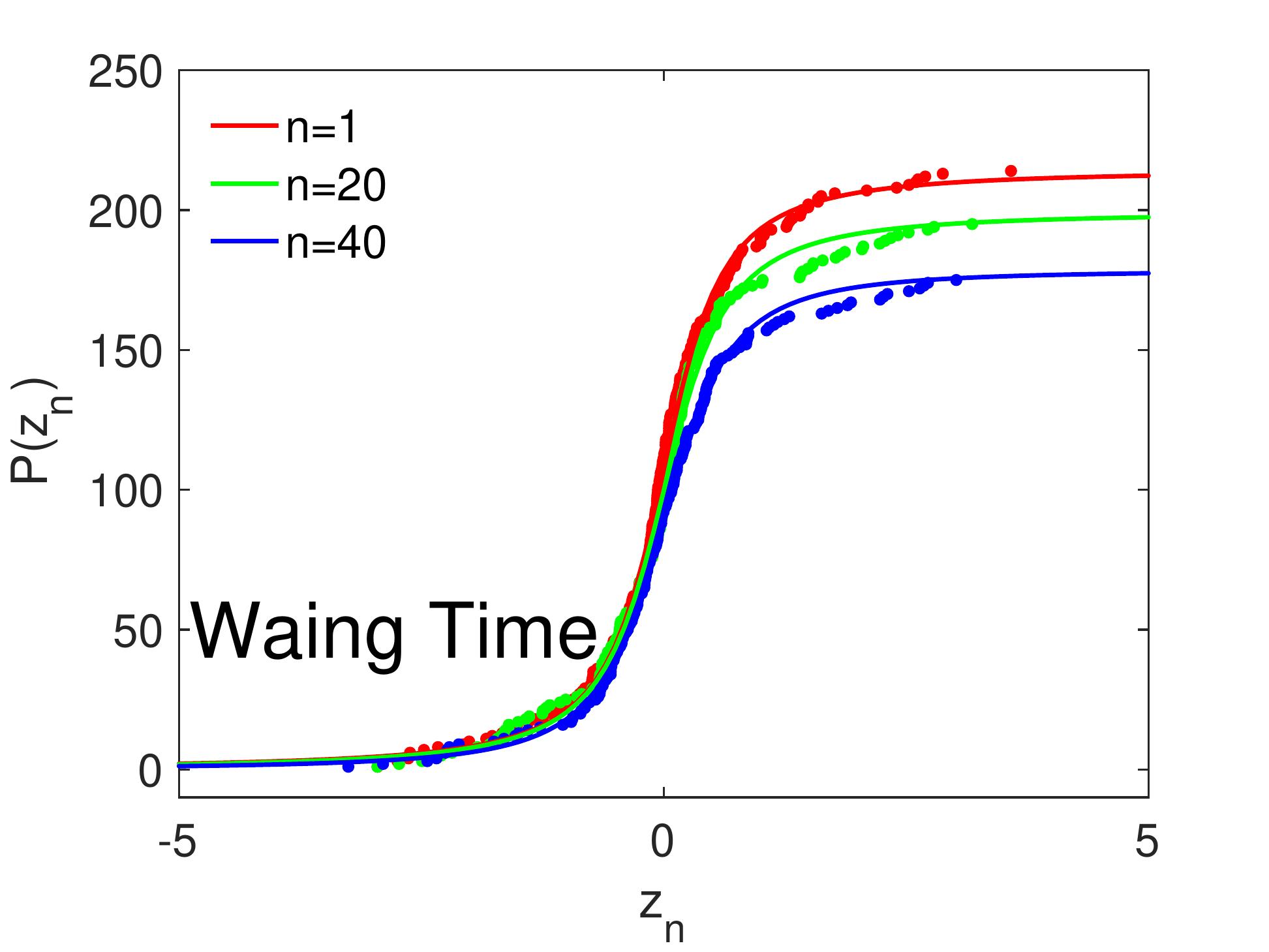}
 \includegraphics[width=0.4\textwidth]{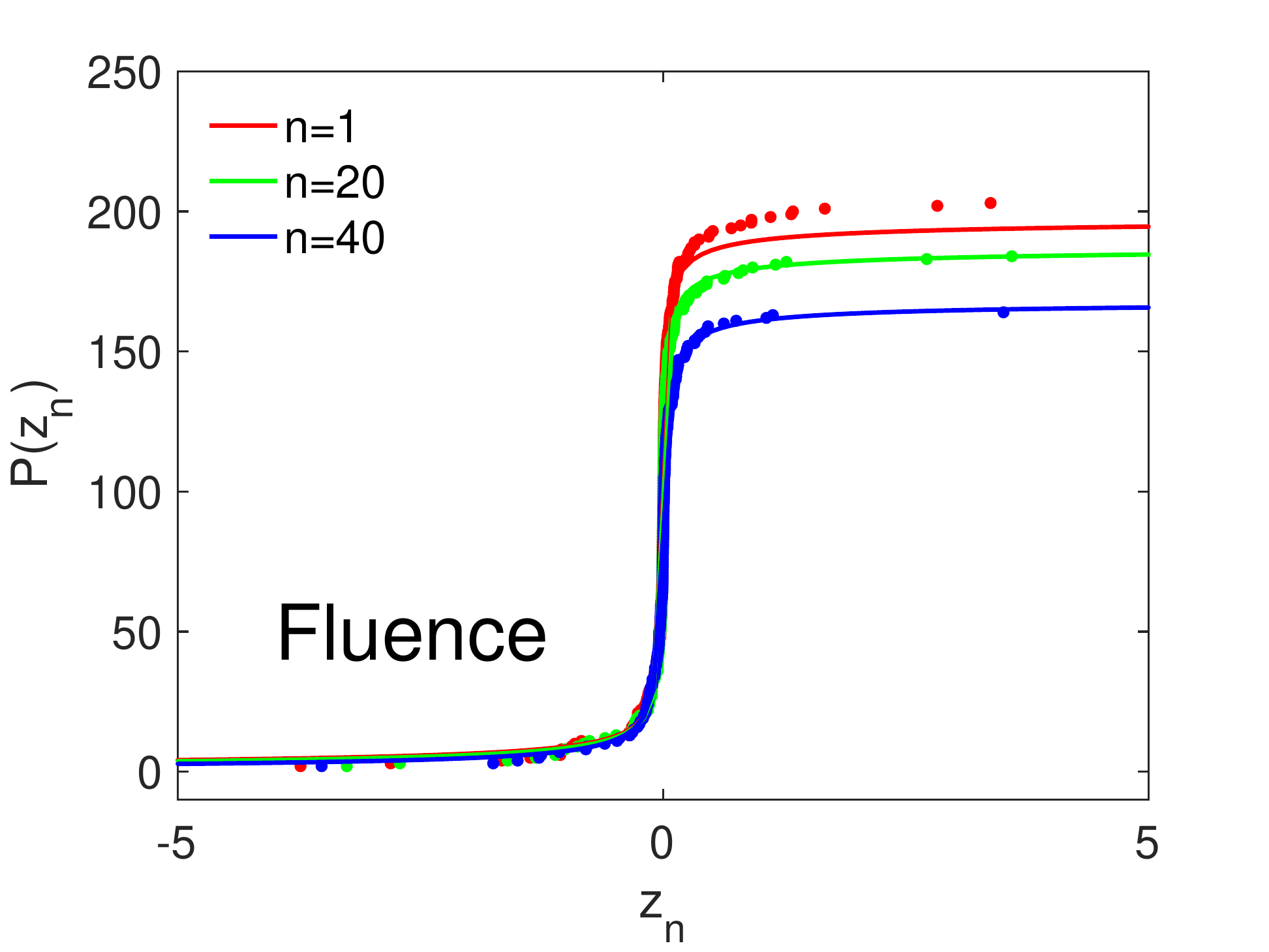}
 \includegraphics[width=0.4\textwidth]{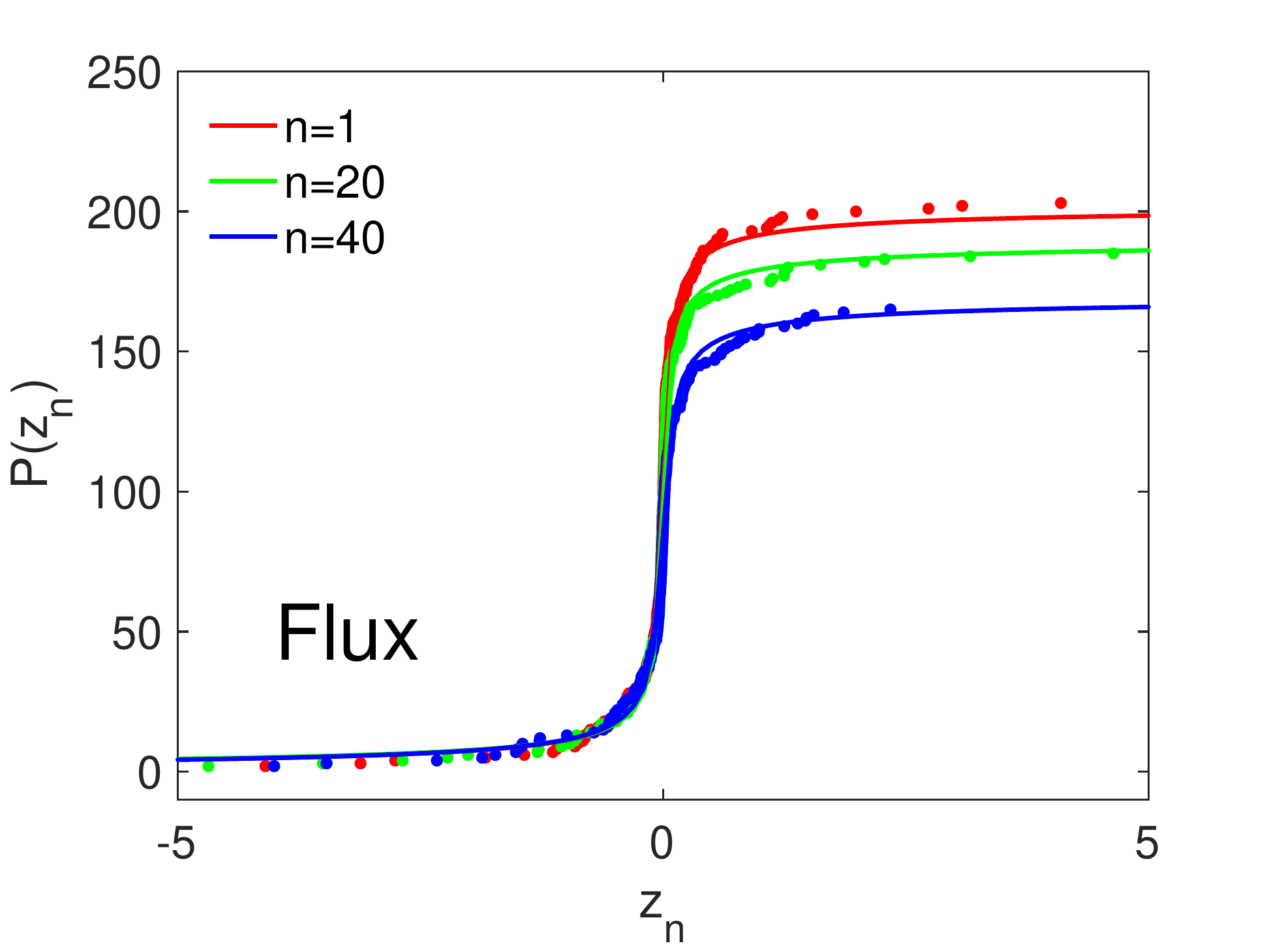}
 \caption{Examples of the CDFs of fluctuations of $T_{90}$, WT, fluence and flux for NICER sample.}
 \label{fig:fluctuation_NICER}
\end{figure}

The CDFs of fluctuations of $T_{90}$, WT, fluence and flux in NICER sample are shown in Figure \ref{fig:fluctuation_NICER}. The color dots denote the data points and the color solid lines denote the best-fitting curves to $q$-Gaussian function. The red, green and blue colors show the fluctuations in temporal interval scale with $n = 1,20,40$, respectively. In Figure \ref{fig:fluctuation_NICER}, we do not plot the errorbars to make the data points to be more distinguishable. But the errorbars are used in the fitting procedure. The best-fitting $q$ values and the corresponding $\chi^2_{\rm red}$ values for $n = 1,20,40$ are listed in Table \ref{tab:qValue}. One could see that the $q$-Gaussian function fits the fluctuations of $T_{90}$, WT, fluence and flux very well.

\begin{table}
  \centering
  \caption{The best-fitting $q$ values for $n = 1,20,40$ for the NICER (upper) and GBM (lower) samples. The numbers in the brackets are the $\chi^2_{\rm red}$ values.}
  \label{tab:qValue}
  \begin{tabular}{clcccc}
  \hline
   sample & & T90& WT & Fluence & Flux  \\
  \hline
  \multirow{3}{*}{NICER}
  & $n=1$ & $1.67\pm0.01 (0.10)$ & $1.79\pm0.02 (0.26)$ & $2.35\pm0.01 (0.28)$ & $2.26\pm0.01 (0.36)$ \\
  & $n=20$ & $1.58\pm0.01 (0.08)$ & $1.77\pm0.02 (0.37)$ & $2.33\pm0.01 (0.17)$ & $2.28\pm0.01 (0.30)$ \\
  & $n=40$ & $1.49\pm0.02 (0.14)$ & $1.70\pm0.01 (0.12)$ & $2.25\pm0.01 (0.21)$& $2.24\pm0.01 (0.29)$\\
  \hline
  \multirow{3}{*}{GBM}
  & $n=1$ & $1.94\pm0.02 (0.14)$ & $2.33\pm0.02 (0.29)$ & $2.25\pm0.01 (0.06)$ & \ldots \\
  & $n=20$ & $1.76\pm0.02 (0.03)$ & $2.20\pm0.02 (0.19)$ & $2.14\pm0.02 (0.09)$ & \ldots \\
  & $n=40$ & $1.87\pm0.08 (0.39)$ & $2.24\pm0.03 (0.24)$ & $2.17\pm0.02 (0.08)$ & \ldots \\
  \hline
  \end{tabular}
\end{table}

\begin{figure}
 \centering
 \includegraphics[width=0.4\textwidth]{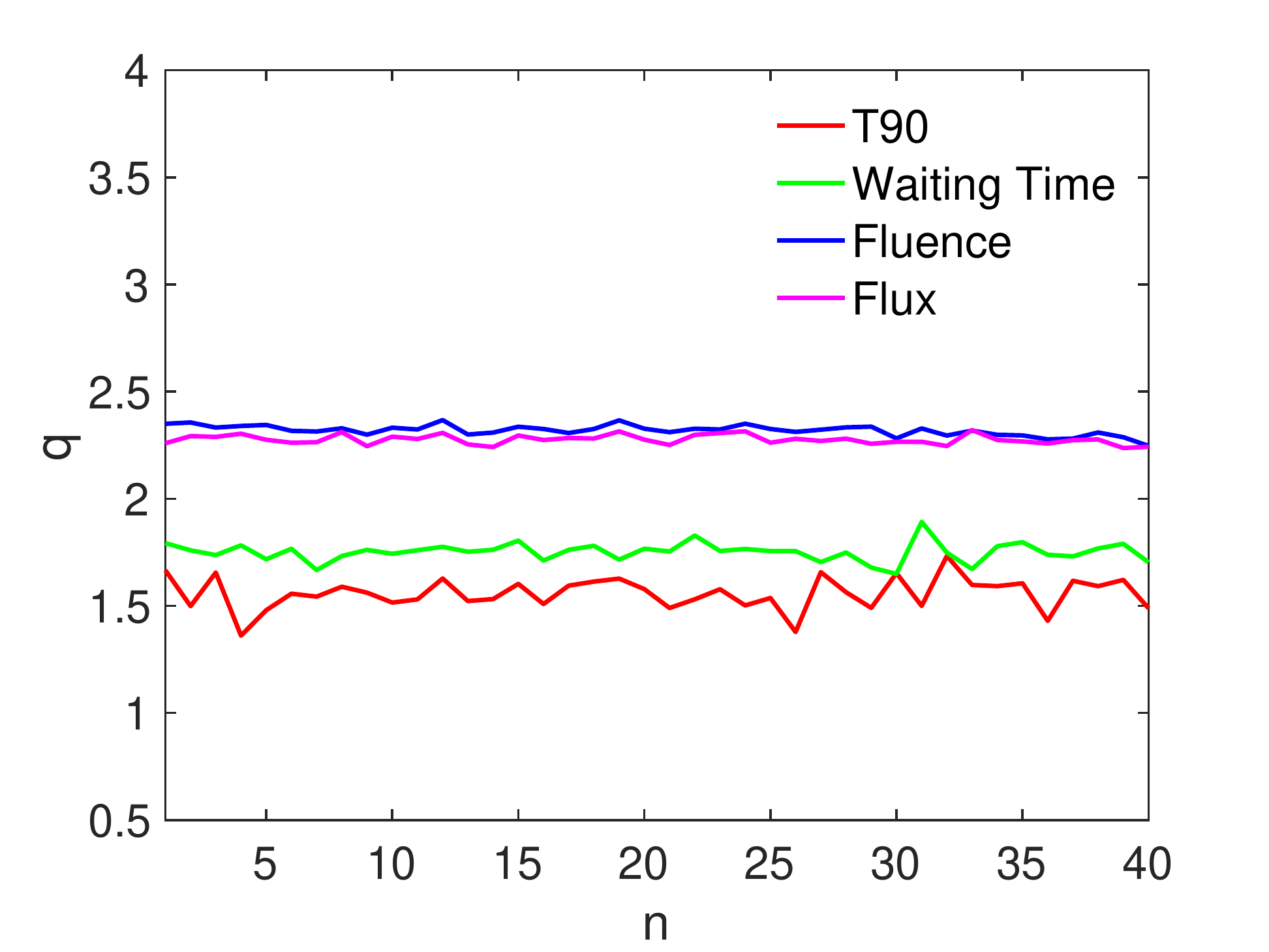}
 \includegraphics[width=0.4\textwidth]{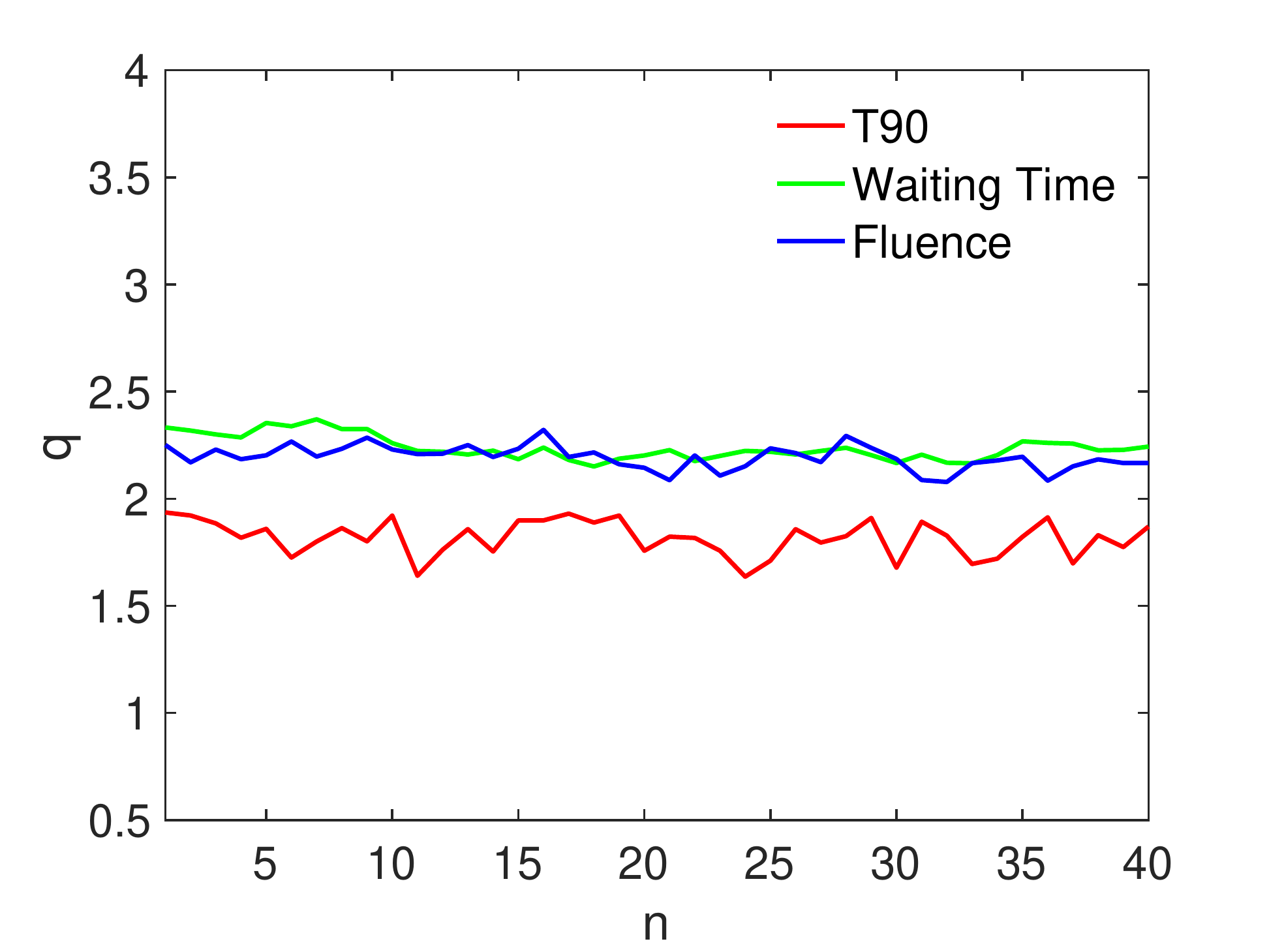}
 \caption{The best-fitting $q$ values for $n = 1 - 40$ for the NICER (left) and GBM (right) samples.}
 \label{fig:qValue}
\end{figure}

\begin{figure}
 \centering
 \includegraphics[width=0.4\textwidth]{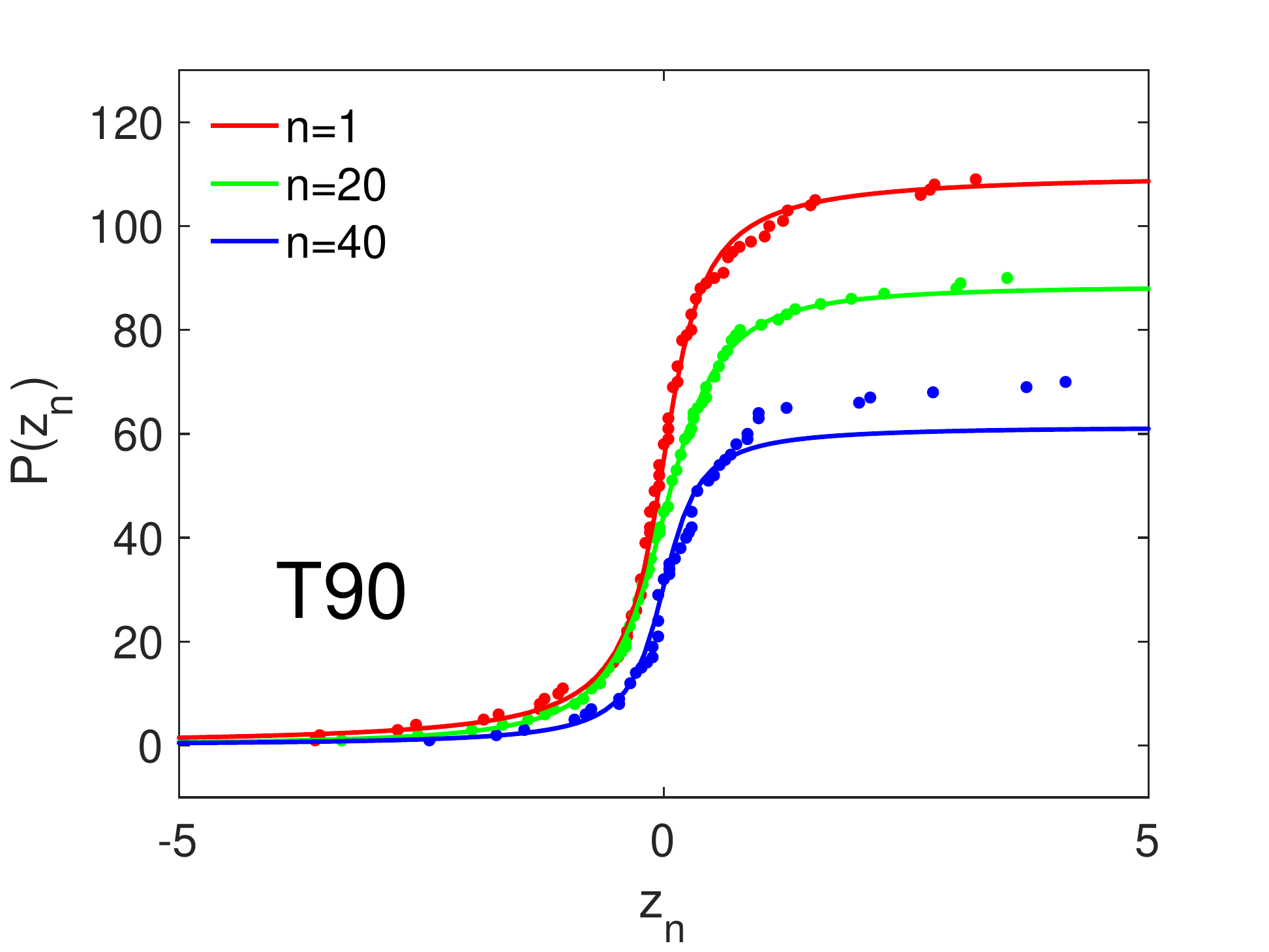}
 \includegraphics[width=0.4\textwidth]{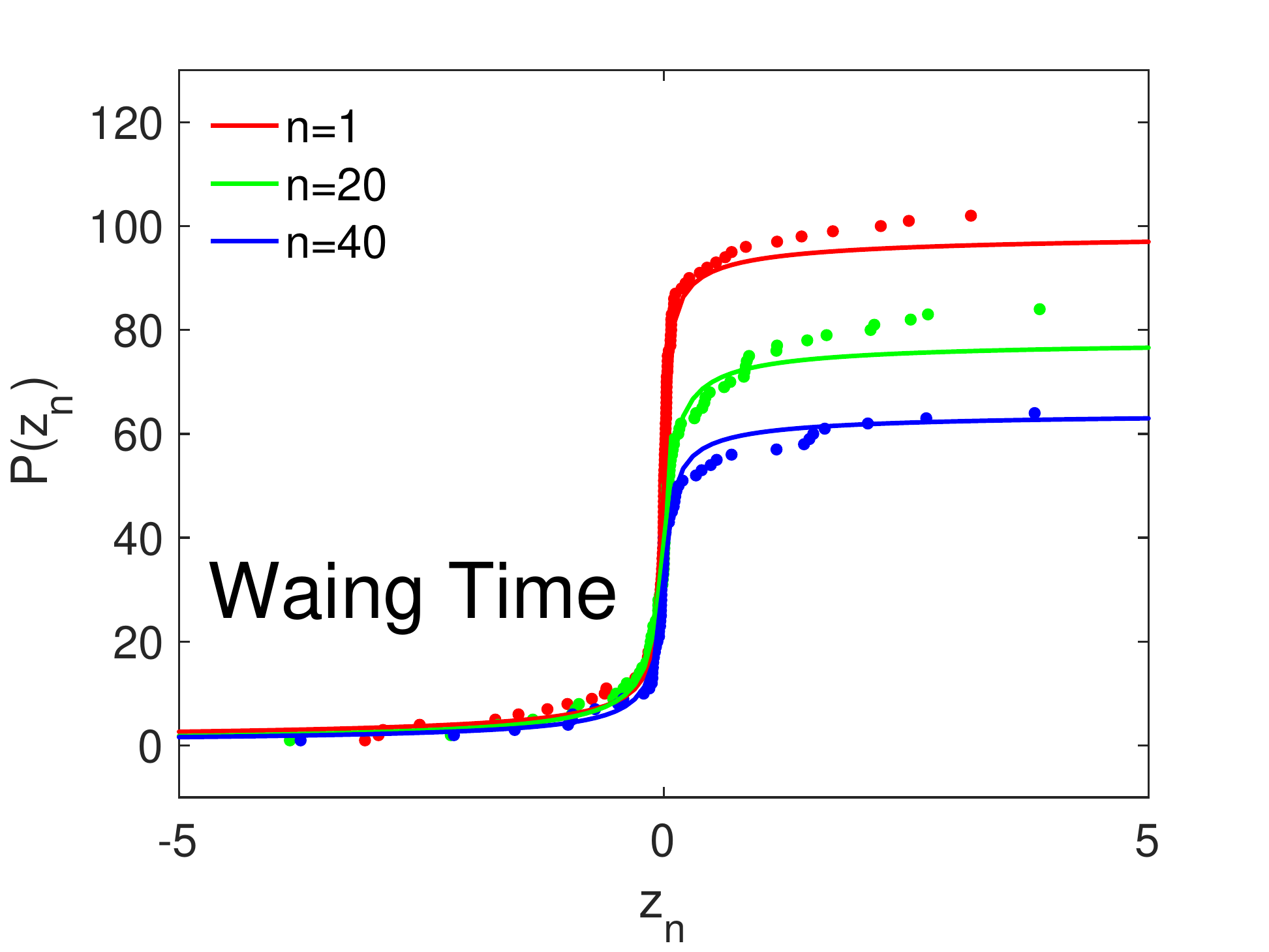}
 \includegraphics[width=0.4\textwidth]{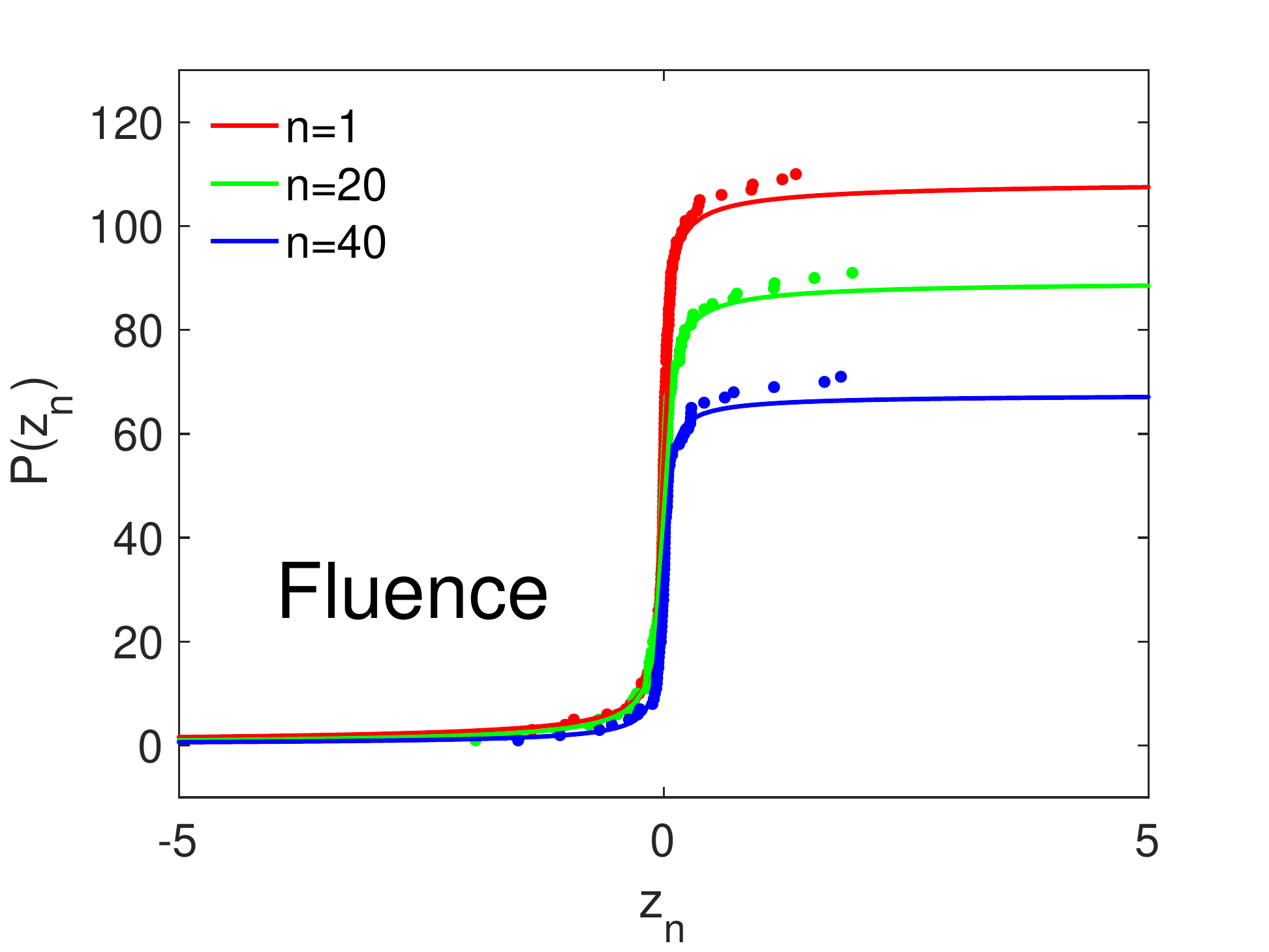}
 \caption{Examples of the CDFs of fluctuations of $T_{90}$, WT and fluence for GBM sample.}
 \label{fig:fluctuation_Fermi}
\end{figure}

We further fit the CDFs of fluctuations in NICER sample for all the temporal interval scale with $n = 1 - 40$. The left panel in Figure \ref{fig:qValue} shows the best-fitting $q$ values as a function of $n$ for $T_{90}$, WT, fluence and flux, respectively. One can see that $q$ values are nearly constant and independent of $n$. The mean $q$ values from  $n = 1 $ to $40$ are $1.56\pm0.08$, $1.75\pm0.04$, $2.32\pm0.02$ and $2.28\pm0.02$ for $T_{90}$, WT, fluence and flux, respectively, with the uncertainty denoting the standard deviation of $q$ values. The consistency of $q$ values for different temporal interval scale indicates the scale invariance of SGRs. This property have also been found in the earthquakes  \citep{Wang:2015nsl} and repeating FRBs \citep{Lin:2019ldn}. This implies that there may be underlying association between SGRs and repeating FRBs.

For the GBM sample, the CDFs of fluctuations of $T_{90}$, WT and fluence are shown in Figure \ref{fig:fluctuation_Fermi}. The best-fitting $q$ values and the corresponding $\chi^2_{\rm red}$ values for $n = 1,20,40$ are listed in the Table \ref{tab:qValue}. The $q$-Gaussian function fits the fluctuations very well. The best-fitting $q$ values as a function of $n$ for $T_{90}$, WT and fluence are shown in the right panel of Figure \ref{fig:qValue}. The mean values of $q$ for $n = 1 - 40$ for $T_{90}$, WT and fluence are $1.82\pm0.08$, $2.24\pm0.06$ and $2.19\pm0.06$, respectively. The constant $q$ values independent of $n$ further confirms the scale invariance of SGRs.

\section{Discussion and Conclusions}\label{sec:conclusions}

In this paper, we investigated the statistical properties of the FRB-associated SGR (SGR 1935+2154) using two samples from different observations. The first sample consists of 217 bursts from the \emph{NICER} observations, and the second sample consists of 112 bursts from \emph{Fermi}/GBM. We show that the CDFs of duration, waiting time, fluence and flux can be well fitted by the BPL model, which is much better than the SPL model. We also found the probability density functions of fluctuations of duration, waiting time, fluence and flux can be fitted by the Tsallis $q$-Gaussian function, with a steady $q$ value independent of the temporal scale. This means that the fluctuations of duration, waiting time, fluence and flux are scale invariant.

The results of this paper are consistent with the statistical properties  of SGR J1550-5418. \citet{Chang:2017bnb} investigated 384 bursts in the three active episodes of SGR J1550-5418, and found that the SGR shows similar behavior to earthquakes. The CDFs of fluence, peak flux and duration of SGR J1550-5418 can also be well fitted by BPL model, and the PDFs of fluctuations of fluence, peak flux and duration of SGR J1550-5418 can be well fitted by a $q$-Gaussian function. Hence the power law distribution of the bursts and the scale invariance of the size fluctuations are the common features of SGRs. Those features imply that the physical origin of SGRs are similar to the earthquakes on earth, such as the crustquakes of neutron stars with extremely strong magnetic fields.

We note that the $q$-values for the NICER and GBM samples are quite different, although the bursts of these two samples come from the same SGR source. The discrepancy may be explained both experimentally and theoretically. In the experimental aspect, the two data samples are observed in different episodes. The GBM sample is observed during four active episodes in 2014, 2015 and 2016, but NICER sample is observed during the first 1120 seconds of the burst storm. In the theoretical aspect, different $q$ values corresponding to different faulting styles. \citet{Wang:2015nsl} also found that different faulting styles have different $q$ values in earthquakes.

Recent discovery of a Galactic FRB 200428 associated with an X-ray burst from the Galactic magnetar SGR J1935+2154 suggests that at least some FRBs originate from magnetars. \citet{Yang:2021thb} proposed that FRBs are triggered by crust fracturing of magnetars, with the burst event rate depending on the magnetic field strength in the crust. This idea is support by the results in this paper and our previous study on repeating FRB 121102 \citep{Lin:2019ldn}. In the previous study, we investigated the statistical properties of the repeating FRB 121102 using two samples from different observations. We showed that the CDF of fluence, flux density, total energy and waiting time can be well fitted by the BPL model. More importantly, the PDFs of fluctuations of fluence, flux density and total energy of the repeating FRB 121102 well follow the Tsallis $q$-Gaussian distribution. The $q$ values keep steady around $q\sim 2$ for different scale intervals, indicating a scale-invariant structure of the bursts. Compared to the repeating FRB 121102 where $q$ values for fluence, flux density and total energy are almost the same ($q=2.04\pm0.05$, $2.14\pm0.05$ and $2.13\pm0.04$, respectively), the $q$ values of $T_{90}$, WT, fluence and flux in SGR J1935+2154 spread a little bit large ($q=1.5\sim 2.4$) and depends on data samples. Nevertheless, the power law distribution and scale invariance of the fluctuations in the repeating FRBs are very similar to the features of SGRs and earthquakes. These studies imply that FRBs, at least some repeating FRBs may originate from the starquakes on a compact star, just like the earthquakes on the Earth (see \citet{Suvorov:2019rzz} for some possible models).

\vspace{2mm}
{\it Note: When the manuscript is in preparation, a similar work appears on arXiv \citep{Wei:2021kdw}. The authors studied the statistical properties of soft gamma-/hard X-ray bursts from SGRs 1806--20 and J1935+2154 and of radio bursts from the repeating FRB 121102. They found similar scale-invariant structure in SGRs and FRBs, which are consistent with our results.}

\section*{Acknowledgements}
This work has been supported by the National Natural Science Foundation of China under Grant Nos. 12005184, 11603005, 11775038 and 12147106.

\section*{Data availability}
The data underlying this article are available in references \citep{Younes:2020hie,Lin:2020mlk}.

\bibliographystyle{mnras}
\bibliography{reference}
\label{lastpage}

\end{document}